\documentclass[12pt]{article}
\usepackage{amsmath,amssymb,amsthm,amsxtra,overpic,bbm,bm,epsfig,subfigure}
\usepackage{hyperref}
\usepackage{mathrsfs}
\usepackage{graphicx}
\usepackage{multirow}
\usepackage{color}
\usepackage{comment}
\usepackage{epstopdf}
\usepackage{float}
\usepackage{cite}
\usepackage{hyperref}
\usepackage{url}
\usepackage{slashed,stmaryrd}
\usepackage{longtable}
\usepackage{extarrows}

\begin{document}
	\def\thefootnote{\fnsymbol{footnote}}
	\vspace{0.2cm}
	\begin{center}
		{\Large\bf Reply to ``Comment on `Flavor invariants and renormalization-group equations in the leptonic sector with massive Majorana neutrinos' ''}
\end{center}

\vspace{0.2cm}
	
\begin{center}
		{\bf Yilin Wang}~$^{a,~b}$~\footnote{E-mail: wangyilin@ihep.ac.cn}
		\quad
		{\bf Bingrong Yu}~$^{a,~b}$~\footnote{E-mail: yubr@ihep.ac.cn}
		\quad
		{\bf Shun Zhou}~$^{a,~b}$~\footnote{E-mail: zhoush@ihep.ac.cn}
		\\
		\vspace{0.2cm}
		{\small
$^a$Institute of High Energy Physics, Chinese Academy of Sciences, Beijing 100049, China\\
$^b$School of Physical Sciences, University of Chinese Academy of Sciences, Beijing 100049, China}
\end{center}
	
\vspace{0.5cm}
	
\begin{abstract}
In the preprint arXiv:2110.08210, some comments on our paper recently published in JHEP \textbf{09} (2021) 053 have been made. Since some of the comments are completely wrong and others are quite misleading, we decide to clarify the relevant issues in a formal reply. First of all, nothing is wrong about our calculations and physical results in the original paper. Second, there is no logical gap to fill at all. The fact that the $N$-dimensional unitary group over the field of complex numbers ${\rm U}(N,\mathbb{C})$ is not a linear algebraic group is actually irrelevant for the validity of the Molien-Weyl formula. As we shall explain in this reply, all the comments in arXiv:2110.08210 arise from the misunderstanding and misinterpretation of our discussions and results.
\end{abstract}
	
	\newpage
	
	\def\thefootnote{\arabic{footnote}}
	\setcounter{footnote}{0}
\framebox{\bf 1} --- To begin with, we briefly describe the theoretical framework and summarize the main results in our paper~\cite{Wang2021}.
At the low-energy scale, the effective Lagrangian for lepton masses, flavor mixing and CP violation in the leptonic sector with three massive Majorana neutrinos is given by
\begin{eqnarray}
	\label{eq:effective lagrangian}
	{\cal L}^{}_{\rm lepton} = - \overline{l^{}_{\rm L}} M^{}_l l^{}_{\rm R} - \frac{1}{2} \overline{\nu^{}_{\rm L}} M^{}_\nu \nu^{\rm C}_{\rm L} + \frac{g}{\sqrt{2}} \overline{l^{}_{\rm L}} \gamma^\mu \nu^{}_{\rm L} W^-_\mu + {\rm h.c.} \; ,
\end{eqnarray}
where $\nu_{\rm L}^{\rm C} \equiv {\cal C}\overline{\nu_{\rm L}^{}}_{}^{\rm T}$ with ${\cal C}\equiv {\rm i} \gamma^2 \gamma^0$ being the charge-conjugation matrix, while $M_l^{}$ and $M_\nu^{}$ denote the charged-lepton mass matrix and the Majorana neutrino mass matrix, respectively. It is obvious that the $3\times 3$ mass matrices $M_l^{}$ and $M_\nu^{}$ will be changed by the flavor-basis transformations $M^{}_l \to U^{}_{\rm L} M^{}_l U^\dagger_{\rm R}$ and $M^{}_\nu \to U^{}_{\rm L} M^{}_\nu U^{\rm T}_{\rm L}$, where $U^{}_{\rm L}$ and $U^{}_{\rm R}$ are arbitrary $3\times 3$ unitary matrices in the flavor space. However, the physical observables, such as the lepton masses, flavor mixing angles and CP-violating phases, are independent of the chosen flavor basis and thus should be related to the so-called flavor invariants.

By flavor invariants, we actually mean the matrix invariants with the building blocks $M^{}_l$ and $M^{}_\nu$ under the aforementioned unitary transformations. The exact definition of flavor invariants has been explicitly given in our paper~\cite{Wang2021}
\begin{eqnarray}
	\label{eq:general form of invariants}
	I^{abcd\cdots}_{rstu\cdots}\equiv {\rm Tr}\left\{H_l^a H_{\nu}^b G_{l\nu}^c \left[G_{l\nu}^{(n)}\right]^d H_l^r H_{\nu}^s G_{l\nu}^t \left[G_{l\nu}^{(n^{\prime})}\right]^u \cdots\right\}\;,
\end{eqnarray}
where $H_l^{}\equiv M_l^{}M_l^{\dagger}, H_\nu^{}\equiv M_\nu^{}M_\nu^{\dagger}, G_{l\nu}^{}\equiv M_\nu^{}H_l^* M_\nu^{\dagger}, G_{l\nu}^{(n)}\equiv M_\nu^{}(H_l^*)_{}^n M_\nu^{\dagger}$ with $n\geq 2$ being integers and the non-negative integers $\left\{a, b, c, d, r, s, t, u\right\}$ stand for the power indices of the corresponding Hermitian matrices. By using the Molien-Weyl formula in the plethystic program, we calculate the Hilbert series of the invariant ring, where the symmetry group is ${\rm U}(3,\mathbb{C})$ and lepton mass matrices have been assigned into certain representations of the ${\rm U}(3,\mathbb{C})$ group. We have found 34 basic flavor invariants with the building blocks $M_l^{}$ and $M_\nu^{}$ in the generating set, and any flavor invariant can be decomposed into the polynomial of these 34 basic invariants. The explicit constructions of 34 basic flavor invariants have been performed, and the systematic method of the decomposition and the construction of the polynomial identities among the basic invariants (i.e., the syzygies) at certain degrees are also discussed in Ref.~\cite{Wang2021}.

\vspace{0.3cm}

\framebox{\bf 2} --- The author of the preprint arXiv:2110.08210~\cite{Lu2021} criticizes our study by arguing that ${\rm U}(N,\mathbb{C})$ is not a reductive group and thus it is unjustified to directly use the Molien-Weyl formula to calculate the Hilbert series of the invariant ring for the ${\rm U}(3,\mathbb{C})$ group. In addition, the author of Ref.~\cite{Lu2021} makes some misleading remarks on the relationship between the physical observables and the flavor invariants. In the following discussions, we shall explain why both the physical remarks and the mathematical statements made in Ref.~\cite{Lu2021} are wrong and all the physical results in Ref.~\cite{Wang2021} are correct.
\begin{itemize}
\item In Ref.~\cite{Wang2021}, we have reached the conclusion that all the flavor invariants defined in Eq.~(\ref{eq:general form of invariants}) can be expressed as the polynomials of those 34 basic invariants in the generating set. The author of Ref.~\cite{Lu2021} has misinterpreted our conclusion in the way that the Jarlskog invariant should be written as a polynomial of 34 basic invariants. This is completely wrong and very misleading. In 1985, Jarlskog constructed the first flavor invariant in the quark sector, which can be extended to the leptonic sector as
    \begin{eqnarray}
    {\rm Det} \left[H^{}_l, H^{}_\nu\right] = 2 {\rm i} \Delta^{}_{e\mu} \Delta^{}_{\mu\tau} \Delta^{}_{\tau e} \Delta^{}_{12} \Delta^{}_{23} \Delta^{}_{31} {\cal J} \; ,
    \label{eq:jarlskog}
    \end{eqnarray}
    where $\Delta^{}_{\alpha \beta} \equiv m^2_\alpha - m^2_\beta$ (for $\alpha\beta = e\mu, \mu\tau, \tau e$) and $\Delta^{}_{ij} \equiv m^2_i - m^2_j$ (for $ij = 12, 23, 31$) denote respectively the charged-lepton and neutrino mass-squared differences. The Jarlskog invariant, which is normally referred to as an invariant under redefining the phases of fermion fields, can be written as ${\cal J} \equiv \sin\theta^{}_{12} \cos \theta^{}_{12} \sin\theta^{}_{13} \cos^2\theta^{}_{13} \sin \theta^{}_{23} \cos\theta^{}_{23} \sin \delta$ in terms of three flavor mixing angles $\{\theta^{}_{12}, \theta^{}_{13}, \theta^{}_{23}\}$ and the Dirac-type CP-violating phase $\delta$ in the standard parametrization of the flavor mixing matrix. Obviously, the determinant as a whole in Eq.~(\ref{eq:jarlskog}) is proportional to ${\rm Tr}\left([H^{}_l, H^{}_\nu]^3\right)$, which is a flavor invariant by definition in Eq.~(\ref{eq:general form of invariants}),\footnote{The determinant in Eq.~(\ref{eq:jarlskog}) is actually one of the basic invariants listed in Ref.~\cite{Wang2021}, namely, $I_{25}^{}\equiv {\rm Tr}\left(H_l^2 H_\nu^2 H_l^{} H_\nu\right)-{\rm Tr}\left(H_l^2 H_\nu^{} H_l^{} H_\nu^2\right)$.} but the Jarlskog (rephasing) invariant ${\cal J}$ itself is not. Similar arguments apply as well to other physical observables, such as lepton masses, flavor mixing angles and CP-violating phases. It is disappointing that the author of Ref.~\cite{Lu2021} doesn't quite understand the relationship between the physical observables and the flavor invariants, which has been clearly stated in the original paper~\cite{Wang2021}. The physical observables are certainly not affected by the flavor-basis transformations, and they are indeed functions, but not polynomial functions, of the basic flavor invariants in the generating set.
    
\item The author of Ref.~\cite{Lu2021} has given a critical argument that ${\rm U}(N, \mathbb{C})$ is not a reductive group and thus the calculation of Hilbert series by virtue of Molien-Weyl formula is invalid. However, this mathematical statement is wrong. Although ${\rm U}(N, \mathbb{C})$ is not an algebraic group by definition, there is an isomorphism between the compact Lie group ${\rm U}(N, \mathbb{C})$ to the general linear group ${\rm GL}(N,\mathbb{C})$, which is a reductive algebraic group.  In fact, every compact connected Lie group has a \emph{complexification} that corresponds to a complex reductive algebraic group, and such a construction gives an intimate correspondence between compact connected Lie groups and complex reductive algebraic groups, up to isomorphisms~\cite{them1,them2,them3}. In our case, since any finite-dimensional faithful representations of the compact subgroups of ${\rm GL}(N, \mathbb{C})$ are completely reducible, it has been explicitly stated as a corollary in Ref.~\cite{book1} that every compact subgroup of ${\rm GL}(N, \mathbb{C})$, of course including ${\rm U}(N,\mathbb{C})$, is reductive. Furthermore, one can establish a one-to-one correspondence between the faithful representations of ${\rm U}(N,\mathbb{C})$ and those of ${\rm GL}(N, \mathbb{C})$, which is algebraic and reductive, thus the finiteness theorem and the validity of the Molien-Weyl formula are well justified~\cite{them1,book2}. Therefore, our calculations of the Hilbert series and the physical results stand correct as they do in the original form. 
\end{itemize}

\framebox{\bf 3} --- In summary, Ref.~\cite{Lu2021} just collects a series of definitions and theorems that have been well known in the mathematical literature. The author of Ref.~\cite{Lu2021} does neither understand the relationship between the physical observables and the flavor invariants, which has been clearly clarified in Ref.~\cite{Wang2021}, nor have a comprehensive knowledge about the algebraic groups and invariant theory. Both physical remarks and mathematical statements made in Ref.~\cite{Lu2021} are wrong and none of the physical results in the original paper~\cite{Wang2021} needs to be changed.

\end{document}